\documentclass[%
reprint,
superscriptaddress,
amsmath,amssymb,
aps,
pra,
floatfix,
]{revtex4-2}

\usepackage{graphicx}% Include figure files
\usepackage{dcolumn}% Align table columns on decimal point
\usepackage{bm}% bold math
\usepackage{comment}
\usepackage{color} 
\usepackage{url}

\hypersetup{breaklinks=true}

\begin{document}
%\hfill {Blue:~modification~candidates~(to be discussed)} \hfill

\preprint{APS/123-QED}

\title{Measurement of cesium  $8\,^2P_{J}\rightarrow 6\,^2P_{J'}$ electric
quadrupole transition probabilities\\using fluorescence spectroscopy}
% Force line breaks with \\

%\thanks{A footnote to the article title}%

\author{Jing Wang}
\affiliation{Graduate School of Natural Science and Technology, Okayama University, Okayama 700-8530, Japan}%
%Lines break automatically or can be forced with \\
\author{Yuki Miyamoto}
\affiliation{Research Institute for Interdisciplinary Science,
           Okayama University, Okayama 700-8530, Japan}
\author{Hideaki Hara}
\affiliation{Research Institute for Interdisciplinary Science,
           Okayama University, Okayama 700-8530, Japan}
\author{Minoru Tanaka}
\affiliation{Department of Physics, Osaka University, Toyonaka, Osaka 560-0043, Japan}%
\author{Motomichi Tashiro}
\affiliation{Department of Applied Chemistry, Toyo University, Kujirai 2100, Kawagoe, Saitama 350-8585, Japan}
\author{Noboru Sasao}
\email{sasao@okayama-u.ac.jp}
\affiliation{Research Institute for Interdisciplinary Science,
           Okayama University, Okayama 700-8530, Japan}

\date{\today}% It is always \today, today,
             %  but any date may be explicitly specified

\begin{abstract}
Fluorescence spectra of the $8\,^2P_{J} \rightarrow 6\,^2P_{J'}$ ($J$ and $J'$ = 3/2, 1/2) electric quadrupole transition of cesium atoms have been observed with a heated cesium vapor cell. We determined the ratio of the transition probabilities of $8\,^2P_{J}\rightarrow6\,^2P_{J'}$ to $8\,^2P_{J}\rightarrow5\,^2D_{3/2}$ by comparing their respective photon emission rates. The results are in good agreement with our theoretical calculations. These measurements provide crucial parameters for tests of coherent amplification method and improve knowledge of cesium properties which are essential to dark matter detection through atomic transitions.
\end{abstract}

%\keywords{Suggested keywords}%Use showkeys class option if keyword
                              %display desired
\maketitle

%\tableofcontents
%-----Section 1-----%
\section{Introduction}
\label{sec:introduction}
%-----Section 1-----%
Revolutionary developments in the atomic, molecular, and optical physics have been made in the past few decades \cite{drake2023springer}. Breakthroughs in experimental techniques, including laser cooling and trapping of atoms, buffer gas cooling for molecules, made it possible to perform precise tests of fundamental physics and observe new physics beyond the standard model  \cite{Ginges04, Roberts15, new-physics}. For instance, spectroscopic analyses of cesium (Cs) atom provided the most precise low-energy tests of parity non-conservation in the electroweak interaction \cite{woodPNC,Bouchiat82,Guena05}; the determination of the permanent electric dipole moment, a key indicator of violations of time-reversal and parity symmetries, has been accomplished \cite{CsEDM69, Ensberg67, chin2001,graner2016edm,acme2018edm}or is planned \cite{Inoue15, Weiss03} for various atoms and molecules. {The cornerstone of these measurements is the determination of the atomic or molecular parameters,  including lifetimes, hyperfine structures, and transition matrix elements.}

 Another compelling driver is the dark matter detection.  Atomic or molecular transitions induced by the absorption of light dark matter particles, such as Axions or dark photons, has come up in recent years \cite{sikivie,arvanitaki2018}, and several new experimental methods have been proposed \cite{santamaria2015axion,huang2019}. However, due to the exceedingly low transition probability, an amplification method is needed. Coherent amplification of rare processes, as proposed \cite{Sasao2018} and examined by our group \cite{miyamoto-1,miyamoto-2}, emerges as a vital tool. This method holds the potential to significantly enhance processes like electric quadrupole transitions or two-photon transitions, thereby enabling the study related to extremely weak phenomena, including dark matter absorption and neutrino mass spectroscopy \cite{masuda2016span}.
 
 {
We plan to do a dark matter search experiment using Cs atoms \cite{Wang-Axion}: transitions induced by dark matter from $8\,^2P_{3/2}$ to $6\,^2P_{3/2}$ via $7\,^2D_{3/2}$ would be enhanced by coherence between $8\,^2P_{3/2}$ and $6\,^2P_{3/2}$. We select Cs atoms as the target due to their simple electronic structure and well-studied atomic properties \cite{pucher5dlifetime, quirk8phyper, bayram8phyper, williams7phyper, herd6dfre, liu20008pfre, chan6s5de2, chen6s6de2, tojo6s5de2, lineratio}. To determine the degree of coherence experimentally, we compare the coherence-amplified electric quadrupole transition $8\,^2P_{3/2} \to 6\,^2P_{3/2}$ rate with theoretical expectations; this requires knowledge of the transition matrix element in advance.}

 {
In this article, we present a detailed report of the measurement of Cs $8\,^2P_J\rightarrow6\,^2P_{J'}$ electric quadrupole (E2) transition (Magnetic dipole transition is much weaker than E2 transition because of the selection rules \cite{bransden2003physics}).} We performed laser-induced fluorescence spectroscopy of the transition within a heated vapor cell. We determined the ratio of the E2 transition rate to that of an electric dipole (E1) transition $8\,^2P_{J}\rightarrow5\,^2D_{3/2}$ from the same excited state. Additionally, we calculated the $8\,^2P_{J}\rightarrow6\,^2P_{J'}$ transition probabilities using configuration-interaction many-body perturbation theory (CI+MBPT) and compared them with experimental values. The agreement between theory and experiment is found to be good. This inherently weak transition serves as a crucial reference point for evaluating amplification method and understanding potential background effects arising from other forbidden transitions.

The paper is organized as follows: in Sec. \ref{sec: experiment}, we present our experimental method and setup; in Sec. \ref{sec: data analysis} we discuss data analysis and error budget; in Sec. \ref{sec: result and discussions}, we show the results and compare them with theories; and finally, we present our
conclusions in Sec. \ref{sec:summary}.

%-----Section 2-----%
\section{Experiment}
\label{sec: experiment}
%-----Section 2-----%
\begin{figure}[t]
    \includegraphics[width=0.8\columnwidth]{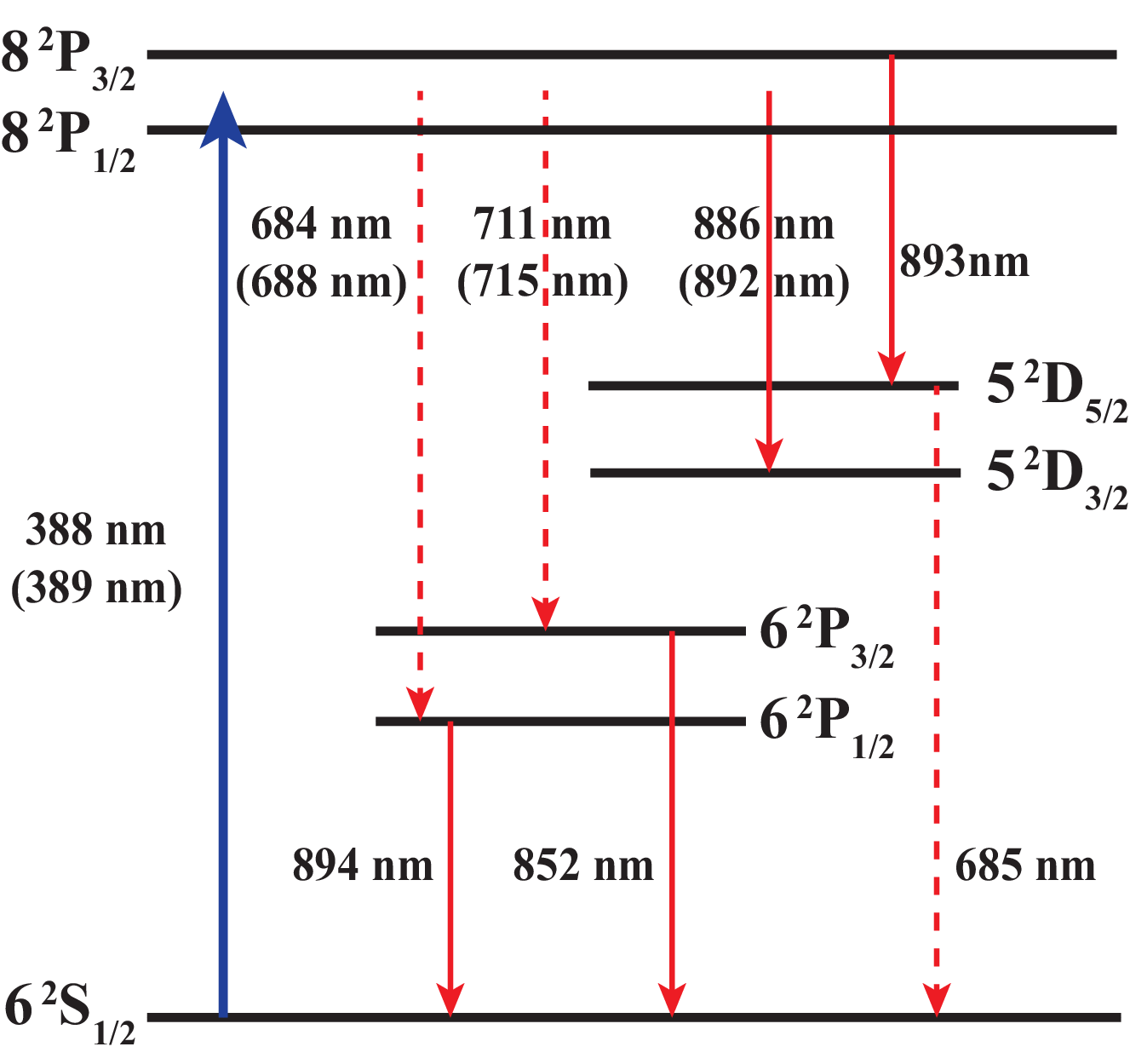}
    \caption{\label{fig:energy level}Energy level of $ {\text{Cs}}$ and relevant transitions in the measurement. Electric dipole (E1) transitions are expressed with solid lines and quadrupole (E2) transitions with dashed lines. { The arrows starting (ending) between the $8\,^2P_{3/2}$ and $8\,^2P_{1/2}$ indicates that they are from (to) one of these states, and their wavelengths for $8\,^2P_{3/2}\;(8\,^2P_{1/2})$ are shown without (inside) parentheses. }We excite the Cs atom from $6\,^2S_{1/2}$ to $8\,^2P_{J}\,(J=3/2,1/2)$ state, and determine the ratio of the transition probabilities $8\,^2P_J\rightarrow6\,^2P_{J'}$ to $8\,^2P_J\rightarrow5\,^2D_{3/2}$ by comparing their photon emission rates.}
\end{figure}
\subsection{Atomic energy level}
The partial energy-level diagram and the relevant transitions of Cs are illustrated in Fig.~\ref{fig:energy level}. 
In the experiment, a continuous-wave laser in resonance with the transition $6\,^2S_{1/2}\rightarrow8\,^2P_{J}$ ($J=3/2,1/2$) at 388, 389 nm is used to excite atoms to the $8\,^2P_{J}$ state.
Successful excitation is detected by monitoring the Cs $D_2$ line ($6\,^2P_{3/2}\rightarrow6\,^2S_{1/2}$) at 852 nm. 
The photons emitted through the E2 transitions $8\,^2P_{J}\rightarrow6\,^2P_{J'}$ ($J'=3/2,1/2$) are then detected except for $8\,^2P_{1/2}\rightarrow6\,^2P_{1/2}$. Additionally, the E1 transitions $8\,^2P_{J}\rightarrow5\,^2D_{3/2}$, are detected as convenient references for determining the ratio of the transition probabilities.
There are several reasons to choose the $8\,^2P_{J}\rightarrow5\,^2D_{3/2}$ transitions as references: 
(i) They originate from the same excited state, ensuring that the ratio is unaffected by potential errors arising from backgrounds such as photo-ionization or multi-photon transition processes. (ii) We expect a reduced sensitivity to radiation-trapping effects. (iii) Our detection system has higher efficiency at the wavelength of this transition than other lines. 

The energy levels for the $8\,^2P_{1/2}$ are similar to the case of $8\,^2P_{3/2}$. There are several differences, however, in addition to the transition wavelengths: 
{(i) Transitions $8\,^2P_{1/2}\rightarrow5\,^2D_{5/2}$ and $8\,^2P_{1/2}\rightarrow6\,^2P_{1/2}$ do not satisfy the E1 and E2 selection rules \cite{bransden2003physics}, so they are outside of our intended sensitivity for measuring E2 transitions.}
(ii) The transition strength of $6\,^2S_{1/2}\rightarrow8\,^2P_{1/2}$ is smaller than that of $8\,^2P_{3/2}$ by a factor of $\sim 4$ \cite{liu20008pfre}, thus stronger laser intensity is required to have enough signal-to-noise (S/N) ratio, especially for the forbidden transition.

Finally, we note that the initial population distribution of $8\,^2P_{1/2,3/2}$ hyperfine states has no impact on the ratio of transition probabilities we measure since the final hyperfine states are summed up \cite{steck2003cesium}.

\begin{figure}[t]
    \centering
 \includegraphics[width=\columnwidth]{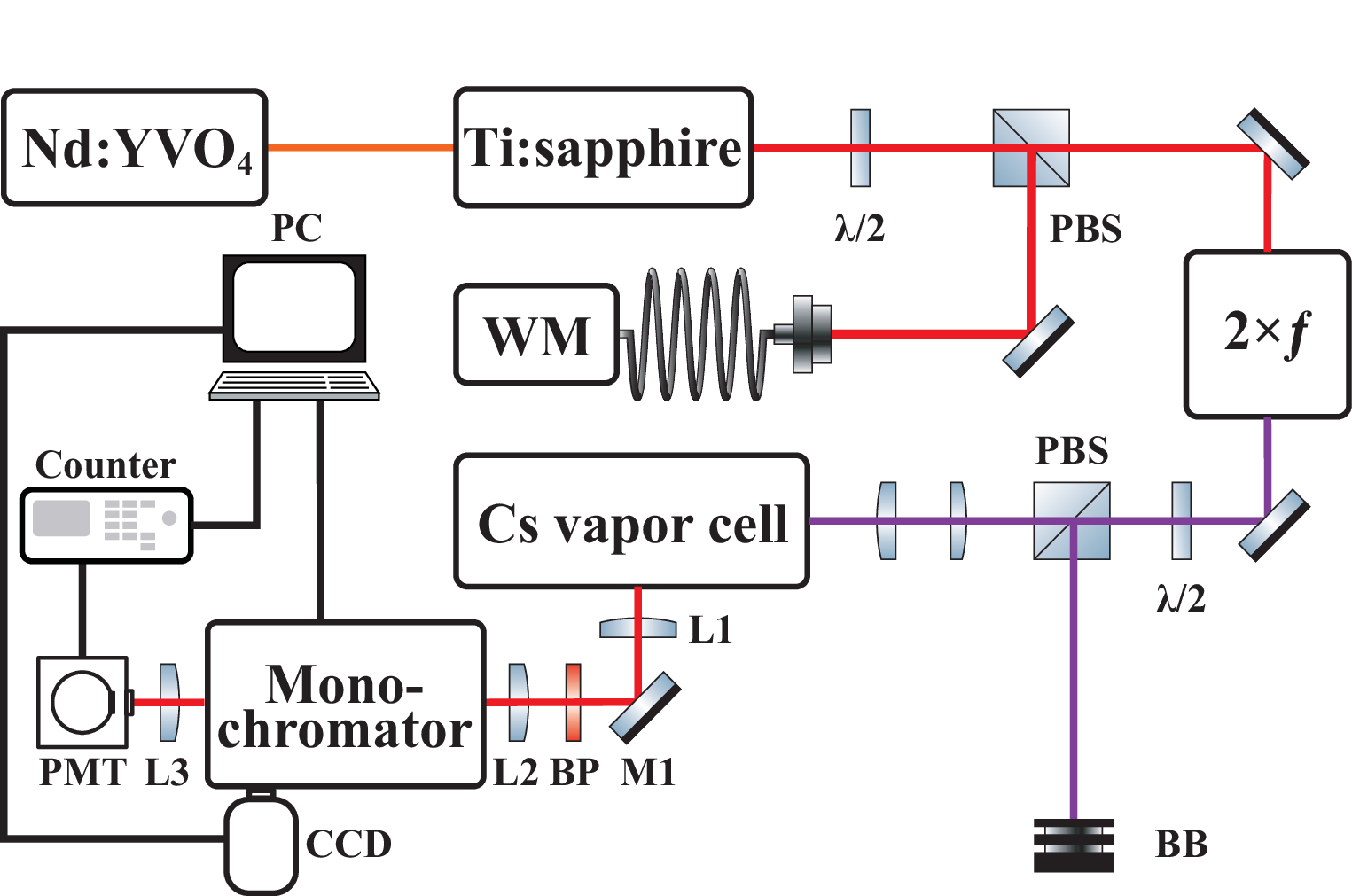}
    \caption{Experimental setup. Nd:YVO$_4$ : neodymium doped yttrium orthovanadate laser, Ti:S: titanium sapphire laser, WM: wavemeter, 2$\times f$: frequency doubling unit, PBS: polarizing beamsplitter, $\lambda/2$: half-wave plate, BB: beam block, BP: band pass filter (see Table \ref{tab:scanned-wavelength}), CCD: charge-coupled device, PMT: photomultiplier tube, L and M: lens and mirror.}
    \label{fig:experiment setup}
\end{figure}
\subsection{Laser system}  
Figure~\ref{fig:experiment setup} illustrates the experimental setup. A Ti:S laser (Coherent, 899-21 Ring laser), pumped by a Nd:YVO$_4$ laser (Coherent, Verdi-10), produces an output of 1 W at 776 nm. This laser can be locked to its reference cavity with a frequency fluctuation of less than 10 MHz. A small fraction of this laser output is directed into a wavemeter (HighFinesse, WS-7) for wavelength determination. The main part of the laser beam is directed into a frequency doubling unit (Coherent, MBD-200). The doubling unit comprises a lithium triborate crystal within a bow-tie cavity and efficiently generates a UV beam with a frequency matching the $6\,^2S_{1/2}\rightarrow8\,^2P_{J}$ transition. Under typical operating conditions, one obtains a linearly polarized UV beam with a power up to 60 mW at the exit of the doubling unit. A half-wave plate combined with a polarizing beamsplitter is employed to adjust the laser power. A telescope modifies the beam waist to $\omega_x \times \omega_y =$ 0.9 $\times$ 1.7 mm. The laser then goes through a 2.5 cm cylindrical Cs vapor cell (TRIAD Technologies, TT-CS-20X75-CW). 
The cell is heated by a heater (Thorlabs, GCH25-75) and temperature is maintained by a temperature controller (Thorlabs, TED200C) with a temperature fluctuation within 0.1$^{\circ}$C.

\begin{figure*}[t!h]
      \centering

         \includegraphics[width=0.87\textwidth]{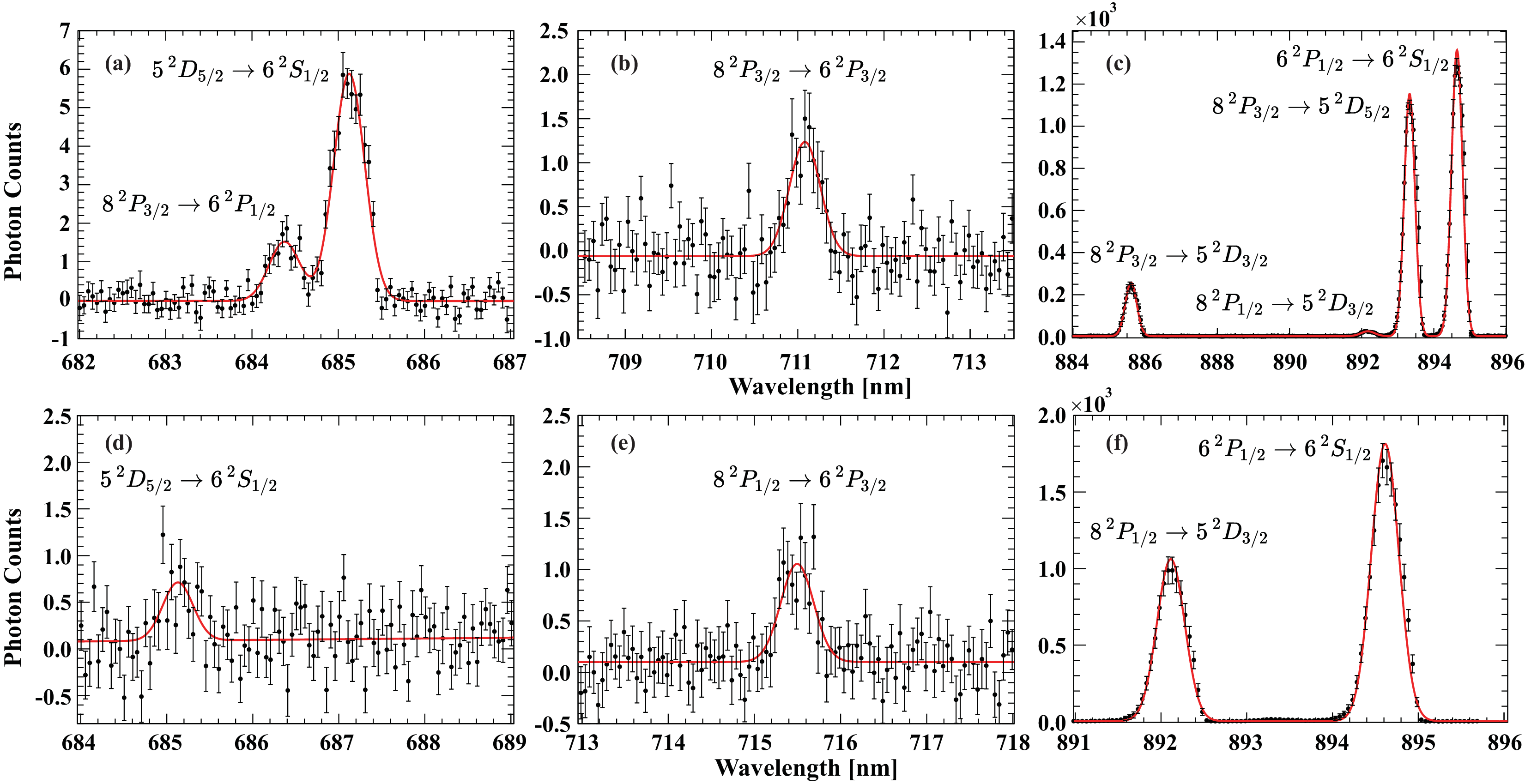}
       
            \caption{Example of raw spectra. Experimental data are shown by black dots and fitting functions by red solid lines. The wavelength bin size is chosen to be 1/20 nm for all spectra.
            Top: (a): E2 transitions $8\,^2P_{3/2}\rightarrow6\,^2P_{1/2}$ and $5\,^2D_{5/2}\rightarrow6\,^2S_{1/2}$, (b): E2 transition $8\,^2P_{3/2}\rightarrow6\,^2P_{3/2}$, (c): E1 transitions  $8\,^2P_{3/2}\rightarrow5\,^2D_{3/2}$, $8\,^2P_{1/2}\rightarrow5\,^2D_{3/2}$, $8\,^2P_{3/2}\rightarrow5\,^2D_{5/2}$, $6\,^2P_{1/2}\rightarrow6\,^2S_{1/2}$. Bottom: (d): E2 transition $5\,^2D_{5/2}\rightarrow6\,^2S_{1/2}$, (e): E2 transition $8\,^2P_{1/2}\rightarrow6\,^2P_{3/2}$, (f): E1 transitions $8\,^2P_{1/2}\rightarrow5\,^2D_{3/2}$,$6\,^2P_{1/2}\rightarrow6\,^2S_{1/2}$.}
        \label{fig:raw sepctra}
\end{figure*}
 
\subsection{Fluorescence detector}
Fluorescence lights emitted at a right angle to the propagation direction of the laser are focused on the entrance slit of a 1200-groove/mm monochromator (Princeton Instruments, Acton SP2300i) through a set of optical components such as mirrors, lenses, and band pass filters. The monochromator has two selectable output ports: one of them with an exit slit is connected to a photomultiplier tube (PMT Hamamatsu, R13456P), and the other with no exit slit to a charge-coupled device (CCD) camera (Princeton Instruments, PIXIS-100). 
Normally PMT is cooled down to $-30^{\circ}$C to reduce dark count rates. It has a multi-alkali cathode with a good quantum efficiency ($\sim\!6\%$) at around 700 nm where the forbidden transitions are located. 
The CCD camera operated at $-70^{\circ}$C has higher quantum efficiency than PMT over a wide range of wavelength regions of interest. While it cannot do photon counting, it serves as a fast and convenient detector for determining various resonance frequencies and their intensities. Outputs from the PMT are directed to a pulse counting system comprising a discriminator with a threshold set at approximately 0.8 photo-electron level and a digital counter (Keysight, 53230A). An online computer is employed to record PMT/CCD data and to manage the monochromator.

\begin{table}[b]
 \centering
 \begin{ruledtabular}
 \begin{tabular}{ccccc}
  Label & Type& Scanned region  & Transitions & Filter range \\
  & & $\lambda$ (nm)& &$\lambda$ (nm)\\
   \hline
\rule{0pt}{3ex}(a)  & E2& {682-687} &$8\,^2P_{3/2}\rightarrow6\,^2P_{1/2}$ &[490, 700]$^a$   \\
  (b)  & E2& {708-714} &$8\,^2P_{3/2}\rightarrow6\,^2P_{3/2}$  &[705, 715]$^b$\\
  (c)  & E1& {884-896} & $8\,^2P_{3/2}\rightarrow5\,^2D_{3/2}$ &-\\ 
   & & & $8\,^2P_{3/2}\rightarrow5\,^2D_{5/2}$ & -\\
  \hline
 (d)  & E2& {684-689} &$8\,^2P_{1/2}\rightarrow6\,^2P_{1/2}$ &  [490, 750]$^c$\\
 (e)  & E2& {713-718} &$8\,^2P_{1/2}\rightarrow6\,^2P_{3/2}$&  [490, 750]$^c$
     \\
    (f)  &E1&  {891-896} & $8\,^2P_{1/2}\rightarrow5\,^2D_{3/2}$ & - \\
 \end{tabular}
 \caption{Wavelength ($\lambda$) regions scanned by this experiment and transitions of interests. The labels correspond to the labels in Fig.\ref{fig:raw sepctra}. Filter range indicates a wavelength region transmitted by filters of a: FEL490 and FES700, b: FBH710-10, c: FEL490 and FES750 (all filters from Thorlabs).}
 \label{tab:scanned-wavelength}
 \end{ruledtabular}
\end{table}
\subsection{Experimental procedure}
The experimental procedure is structured as follows. 
We first adjust the frequency of the Ti:S laser to maximize the $D_2$ line (852nm) yields using CCD. Actually, we lock the Ti:S laser frequency to 386.60765 THz for $8\,^2P_{3/2}$ excitation and 385.36378 THz for $8\,^2P_{1/2}$ excitation\cite{quirk8phyper,liu20008pfre}.
After switching to the PMT, we set the entrance and exit slit widths of the monochromator to 20 $\mu$m and 450 $\mu$m. These parameters are kept unchanged throughout the entire experiment. 
The actual scanning procedure for a given wavelength range is as follows. 
The photon counts are recorded every second (1 Hz) while the monochromator is scanning at the speed of 1 nm/min. 
After the signal scan, the background spectrum is also recorded by detuning the laser frequency to the off-resonance. It turns out that the background rate is less than 0.5 Hz. 
The whole process is repeated 10 cycles, requiring approximately one hour for data collection of each wavelength range. 
The scanning is performed for both forbidden transitions and allowed transitions. 
See Table~\ref{tab:scanned-wavelength} for the actual scan regions.

For convenience, we refer to a set of spectra obtained according to the above procedure as a ``run". 
We actually took five runs, two for $8\,^2P_{3/2}$ and three for $8\,^2P_{1/2}$, over a period of 10 days.

%-----Section 3-----%
\section{Data Analysis}
\label{sec: data analysis}
%-----Section 3-----%

\subsection{\label{sec:sepc-8p3/2}
Raw spectra for $8\,^2P_{3/2}$ measurement}
For the excitation of the $8\,^2P_{3/2}$ state, the UV laser wavelength is set to 388 nm with a power of 20 mW, and the Cs vapor cell temperature is set to be 55$^{\circ}$C. 
As described in the previous section, the basic quantities obtained in the experiment are the photon counts on the PMT for different transitions. 
Figure~\ref{fig:raw sepctra}(a)-(c) show the raw spectra, each spectra being the background-subtracted counts averaged over 10 cycles. Figure~\ref{fig:raw sepctra}(a) is the spectra in the forbidden region around 684 nm. In addition to the expected transition of $8\,^2P_{3/2}\rightarrow6\,^2P_{1/2}$ (left peak), another E2 transition $5\,^2D_{5/2}\rightarrow6\,^2S_{1/2}$ (right peak) is observed. Figure~\ref{fig:raw sepctra}(b) shows the E2 transition of  $8\,^2P_{3/2}\rightarrow6\,^2P_{3/2}$ while Fig.~\ref{fig:raw sepctra}(c) shows the E1 transitions $8\,^2P_{3/2}\rightarrow5\,^2D_{3/2}$, $8\,^2P_{3/2}\rightarrow5\,^2D_{5/2}$, and $D_1$($6\,^2P_{1/2}\rightarrow6\,^2S_{1/2}$). 
Due to the existence of the photo-ionization effect, some population goes to the higher excited state (\textit{e.g.}, $9S_{1/2}$), and then decay to $8\,^2P_{1/2}$ state, so the  $8\,^2P_{1/2}\rightarrow5\,^2D_{3/2}$ transition (892 nm) is also observed. This population leakage is confirmed by the detection of other $9S_{1/2}$ transitions using CCD. But it is rather weak and has no influence to the final ratio. All the peaks are fitted by a Gaussian line shape function with a linear background. 
The raw counts (the area underneath the peak) and its statistical error, denoted as $C(\lambda)\pm \Delta C$ for each transition $\lambda$, is extracted from the fitted parameters.

We note all of the peaks exhibit similar full width at half maximum ($0.42\pm 0.05$ nm), which is limited by the resolution of the monochromator.  
We also note that the center wavelengths of the peaks are shifted by 0.4 nm compared to the theoretical wavelength for all the transitions, which can be attributed to the outdated calibration of the monochromator, but this shift does not influence the transition probabilities, and is calibrated in Fig. \ref{fig:raw sepctra}.

\subsection{Raw spectra for $8\,^2P_{1/2}$ measurement}
For the $8\,^2P_{1/2}$ excitation, the wavelength of the UV laser was set to 389 nm. 
As mentioned before, the transition strength of $6\,^2S_{1/2}\rightarrow8\,^2P_{1/2}$ is smaller, so we increased the UV laser power from 20 mW to 30 mW. Also, the temperature of the Cs vapor cell was increased to 65$^{\circ}$C to increase the atom density. 
The rest parts of the setup remained unchanged.

Figure~\ref{fig:raw sepctra}(d)-(f) show the raw spectra observed by the $8\,^2P_{1/2}$ excitation. As seen in Fig.\ref{fig:raw sepctra}(d), the $8\,^2P_{1/2}\rightarrow6\,^2P_{1/2}$ transition with a wavelength of 688 nm is not observed as expected \cite{bransden2003physics}. 
The small observed peak corresponds to the transition $5\,^2D_{5/2}\rightarrow 6\,^2S_{1/2}$.

\begin{table*}[t]
\centering
\begin{ruledtabular}
\begin{tabular}{ccccccc}
\rule{0mm}{4mm}  &\multicolumn{3}{c}{$C\pm \Delta C$} & {} \\[1mm] \cline{2-4} 
{\rule{0mm}{5mm}$\lambda$ (nm)} & {\rule{0mm}{5mm} Run 1} & {\rule{0mm}{5mm} Run 2} & {\rule{0mm}{5mm} Run 3} & {$\langle C \rangle \pm \Delta \langle C \rangle$} & {$\eta_{qe}$ (\%)} & {${\eta_{opt}}/{\eta_{opt}^{(0)}}$} \\[1.5mm]
\hline
\rule{0pt}{3ex}684 & $(1.39\pm0.12)\times10^1$ & $(1.47\pm0.13)\times10^1$ & {-} & $ ({1.43 \pm 0.09})\times10^1$ & $ {12.2}$ & $ {1.37}$ \\
711 & $(1.10\pm0.16)\times10^1$ & $(1.17\pm0.13)\times10^1$ & {-} & $ ({1.14 \pm 0.10})\times10^1$ & $ {10.6}$ & $ {1.50}$ \\
886 & $(2.01\pm0.07)\times10^3$ & $(2.28\pm0.09)\times10^3$ &{-} & $ ({2.15 \pm 0.06})\times10^3$ & $ {0.194}$ & $ {0.950}$ \\
893 & $(8.49\pm0.17)\times10^3$ & $(1.02\pm0.02)\times10^4$ & {-} & $ ({9.35 \pm 0.13})\times10^3$ & $ {0.100}$ & $ {0.976}$ \\
715 & $(9.88\pm1.94)\times10^0$ & $(1.19\pm0.15)\times10^1$ & $(9.13\pm1.82)\times10^0$ & $ ({1.03 \pm 0.10})\times10^1$ & $ {10.5}$ & $ {1.27}$ \\
892 & $(9.26\pm0.41)\times10^3$ & $(9.47\pm0.38)\times10^3$ & $(8.58\pm0.35)\times10^3$ & $ ({9.11 \pm 0.22})\times10^3$ & $ {0.110}$ & $ {0.972}$ \\
\end{tabular}
\caption{Raw counts ($C$) from different runs along with averaged counts $\langle C \rangle$, PMT quantum efficiency $\eta_{qe}$, and $\eta_{opt}/\eta_{opt}^{(0)}$.}
\label{tab:raw counts}
\end{ruledtabular}
\end{table*}
\subsection{\label{sec:efficiency}
Raw counts and efficiency corrections}
As mentioned, raw counts $C(\lambda)$ for each transition  {with a wavelength $\lambda$} are obtained by fitting a Gaussian function with a linear background. 
{In the case that the fit results are unsatisfactory, {\it i.e.} $\chi^2/n_{\rm dof}>1$ ($n_{\rm dof}$ being the degree of freedom), the errors are enlarged so that $\chi^2/n_{\rm dof}=1$.}
Table \ref{tab:raw counts} summarizes the results of such fits for each run. 
From the Table, it is found that the corresponding counts measured on different days fluctuate more than statistically expected most likely due to environmental temperature variation or laser conditions. 
See Sec.\ref{sec: uncertainty} for more discussions. 
It is confirmed, however, that the ratios of counts, for example, normalized to 886 nm ($P_{3/2}$) or 892 nm ($P_{1/2}$) data, are statistically consistent with each other (within 2$\sigma$). 
We therefore average the observed counts over different runs. 
The results, denoted by $\langle{C}\rangle(\lambda)$,  are shown in Table \ref{tab:raw counts}.
\par
The count $\langle{C}\rangle(\lambda)$ and its corresponding A-coefficient are related to each other via 
\begin{equation}
\langle{C}\rangle(\lambda)=N_{Cs}A(\lambda)\eta_{geo}\eta_{qe}(\lambda)\eta_{opt}(\lambda)t
\label{eq:raw_counting_rate}
\end{equation}
where $N_{Cs}$ is the number of excited Cs atoms, $\eta_{geo}$ the geometrical acceptance, 
$\eta_{qe}$ the quantum efficiency of PMT, $\eta_{opt}$ the product of efficiencies of optical components from the cell to PMT, and $t$ the data taking time at each point. 
Among the experimental variables in Eq.(\ref{eq:raw_counting_rate}), $N_{Cs}$ does not depend on $\lambda$. Similarly $\eta_{geo}$ can be considered $\lambda$ independent: this is confirmed by the studies using a simulation tool (see Sec.\ref{sec: uncertainty} for details).
Treating $N_{Cs}\eta_{geo}$ is $\lambda$ independent, we define $\displaystyle R(\lambda)=\frac{\langle{C}\rangle(\lambda)}{t} \times \frac{\eta^{(0)}_{opt}}{\eta_{qe}(\lambda)\eta_{opt}(\lambda)}$, ($\eta^{(0)}_{opt}\equiv \eta_{opt}(900\;{\rm nm})$) which is directly proportional to $A(\lambda)$. 
Table \ref{tab:raw counts} shows the list of $\eta_{qe}(\lambda)$ and $\eta_{opt}(\lambda)$ normalized to $\eta_{opt}^{(0)}$. 
They are measured by the separate experiments described in the appendices. %~\ref{appendix:qe}. 

\subsection{Systematic Uncertainty}\label{sec: uncertainty}

We now investigate several potential sources of systematic uncertainties (errors) and study their impact on the measurement. 
 Table~\ref{tab: uncertainty} summarizes our error budget. 
They are classified into two categories: one is a random type (upper 3 rows) and the other is a scale error (lower 4 rows). 
The former introduces fluctuation in counts $C(\lambda)$ in addition to the statistical one.
In contrast, $\eta_{qe}$, $\eta_{opt}$ and $\eta_{geo}$ are proportional constants multiplied by the counts $C(\lambda)$ and thus its error may change {$R$, not $C$}. {The monochromator performance difference also becomes sizeable when the wavelength difference is large.}
They are categorized as ``scale" errors and are treated differently from the random type errors.
Below we explain each of them in more detail. 

Laser Power Fluctuations: One of the dominant contributors to experimental uncertainty is the fluctuations in laser power. These fluctuations amount to approximately 1 mW. The associated uncertainty can be quantified by measuring the photon counts at varying laser power levels. We measured the photon counts at 20 mW, 25 mW and 30 mW and it is discerned that a 1 mW fluctuation introduces an uncertainty of 5$\%$ on the PMT counts.

Laser Frequency Drift: The laser system employed, a Ti:S laser, is stabilized using a reference cavity. However, environmental factors such as optical table vibrations and temperature fluctuations can influence the length of the reference cavity, leading to a drift in the laser's frequency. During the course of the experiment, the laser frequency exhibits a drift of approximately 50 MHz. Notably, this frequency drift lies within the Doppler width. Although it has the potential to impact the ratio of hyperfine states in the excited state, analysis of the CCD camera's spectra reveals that the peak intensity remains stable with a variation of only 10$\%$.

Cell Temperature Variations:  Fluctuations in the temperature of the Cs vapor cell present another source of uncertainty which will influence the Cs density, the fluctuation is about 0.1 $^{\circ}$C. We calculated the Cs density change from the Cs vapor pressure data \cite{alcock1984vapour}, revealing a 0.7 $\%$ change in Cs density for every 0.1 $^{\circ}$C change around our experimental temperature. Additionally, we measured photon counts for $8\,^2P_{1/2}\rightarrow6\,^2P_{1/2}$ transition at temperatures of 55 $^{\circ}$C, 60 $^{\circ}$C and 65 $^{\circ}$C, the linear fitting shows a 0.3 $\%$ change for every 0.1 $^{\circ}$C. To be conservative, we set the uncertainty caused by the temperature fluctuation to be 1 $\%$.

{Radiation trapping effect: 
This effect is most serious for the allowed transition, {\it i.e.} $8\,^2P_{J}\rightarrow5\,^2D_{J'}$. 
To estimate the population of the $5\,^2D_{J'}$ state, we conducted rate equation simulations with the known transition probabilities from NIST atomic spectra database \cite{NIST}, and found that it is comparable to that of the $8\,^2P_{J}$ state ($\sim 2$\% of total Cs atoms). With this result and taking the Doppler effect into account, we estimated the mean free path of the photon using the cross section $\sigma(\lambda) = \frac{g_{2}}{g_{1}}\frac{\lambda_0^{2}}{4} A_{21} g_D(\lambda)$, where $g_1$ and $g_2$ are the degeneracies of the $5\,^2D_{J'}$ and $8\,^2P_{J}$ states, $\lambda_0$ is the transition wavelength of $8\,^2P_{J} \rightarrow 5\,^2D_{J'}$, $A_{21}$ is the A-coefficient \cite{safronova}, and $g_D(\lambda)$ is a Gaussian line shape function with a width equal to the Doppler width $\Gamma_D$ \cite{Foot2005}. Considering the worst case of resonant absorption, $g_D(\lambda_0)=\frac{2\sqrt{ln2}}{\Gamma_D\times\sqrt{\pi}}$, we calculate the mean free path to be 48 cm for $8\,^2P_{3/2} \rightarrow 5\,^2D_{5/2}$, 300 cm for $8\,^2P_{3/2} \rightarrow 5\,^2D_{3/2}$, and 24 cm for $8\,^2P_{1/2} \rightarrow 5\,^2D_{3/2}$, which are much longer than the laser beam radius of $0.9 \times 1.7$ mm. We conclude that the trapping effect is negligibly small.}

\begin{table}[b]
 \centering
 \begin{ruledtabular}
 \begin{tabular}{ccc}
  Source & Uncertainty ($\%$) & Remark\\
   \hline
\rule{0pt}{3ex}  Laser Power & 5\\
  Laser Frequency& 10\\
  Cell Temperature&1\\ 
  Subtotal (random) & 11.2 & Added in quadrature \\
  \hline 
   $\eta_{qe}$ &  {5} & \\
   $\eta_{opt}$ & {5} & \\ 
   $\eta_{geo}$ & {2} & \\
{Monochromator} & {8}&\\
    {Subtotal (scale)} &  {10.9} & Added in quadrature \\
 \end{tabular}
 \caption{{Uncertainties on the photon counts, efficiencies and monochromator acceptance}}
 \label{tab: uncertainty}
 \end{ruledtabular}
\end{table}

\begin{table*}[t]
 \centering
 \begin{ruledtabular}
 \begin{tabular}{ccc|cc|cccc}
 && & 
 \multicolumn{2}{c|}{\textbf{Experiment}} & \multicolumn{4}{c}{\textbf{Theory}} \\
   \textbf{Transition} & $\boldsymbol{\lambda}$ (nm) & \textbf{Type}& $R$ (Hz) & Ratio & $A$ (Hz)\footnote{This work.} & Ratio\footnotemark[1] & $A$ (Hz)\footnote{Safronova {\it et.al.} \cite{safronova}.  {A-coefficient is calculated via $\displaystyle \;A=\frac{\omega_{0}^{3}}{3 \pi \epsilon_{0} \hbar c^{3}} \frac{1}{2 J+1}\left|\left\langle J\|e \mathbf{r}\| J^{\prime}\right\rangle\right|^{2} $ with the energy differences given by NIST atomic sepctra database \cite{NIST}.}} & Ratio\footnotemark[2]  \\
 \hline
\rule{0pt}{3ex}$8\,^2P_{3/2}\rightarrow 6\,^2P_{1/2}$ & 684 & E2 &  {$(8.56\pm0.54)\times10^1$} &  {$(7.32\pm0.50\pm 1.14)\times 10^{-5}$} & 2.27 & $6.99\times 10^{-5}$& & \\
 $8\,^2P_{3/2}\rightarrow 6\,^2P_{3/2}$ & 711 & E2 &  {$(7.17\pm0.63)\times10^1$} &  {$(6.13\pm0.56\pm 0.96)\times 10^{-5}$} & 2.21 & $6.83\times 10^{-5}$ & &\\ 
 $8\,^2P_{3/2}\rightarrow 5\,^2D_{3/2}$ & 886 & E1 &  {$(1.17\pm0.03)\times 10^6$} & 1 & $3.20\times 10^4$ & 1 &$4.93\times 10^4$&1\\
 $8\,^2P_{3/2}\rightarrow 5\,^2D_{5/2}$ & 893 & E1 &  {$(9.58\pm0.13)\times 10^6$} &  {$8.19\pm0.24\pm 0.92$} & $3.24\times 10^5$ & 9.86& $4.55\times 10^5$ & 9.22 \\
 \hline
 $8\,^2P_{1/2}\rightarrow 6\,^2P_{3/2}$ & 715 & E2 &  {$(7.72\pm0.75)\times10^1$} &  {$(9.10\pm0.91\pm 1.42)\times 10^{-6}$} & 4.55 & $11.1\times 10^{-6}$ &&\\ 
 $8\,^2P_{1/2}\rightarrow 5\,^2D_{3/2}$ & 892 & E1 &  {$(8.51\pm0.21)\times 10^6$} & 1 & $4.11\times 10^5$ & 1& $5.66\times 10^5$ & 1 \\
 \end{tabular}
 \end{ruledtabular}
 \caption{Table of experimental and theoretical transition rates and their ratio}
 \label{tab:results}
\end{table*}
Uncertainties in $\eta_{qe}$ and $\eta_{opt}$: As previously mentioned, $\eta_{qe}$ and $\eta_{opt}$ are measured in separate experiments. The uncertainties associated with these factors primarily arise from the non-reproducibility of photo-diode spectra due to the relatively weak signal amplitude, resulting in larger background fluctuations compared to those of the PMT spectra. Both uncertainties are estimated to be 5$\%$. %See Sec. \ref{appendix:qe} for details.
See the appendices for details.
\par
Uncertainties in $\eta_{geo}$: 
The geometrical factor $\eta_{geo}$ depends, in principle, on $\lambda$ through the dispersion effects of the lenses L1 and L2, leading to different spot sizes at the monochromator entrance slit. We performed a ray-tracing simulation using LightTools \cite{ligthtools} and compared $\eta_{geo}$ at two different wavelengths of 700 nm and 900 nm. 
Although the difference depends slightly on input geometrical parameters (such as distances between L1 and L2), it is at most 1.5$\%$. Thus the uncertainty is conservatively taken to be 2$\%$.

{
Monochromator performance difference: The monochromator may exhibit different performance across various wavelength ranges. Since the transitions we want to compare are located either around 700 nm or 900 nm, we categorized the peaks into two groups: one around 700 nm and another around 900 nm.  We calculated the weighted averages of the peak widths in these two regions and compared them. It turns out that the relative difference is around 8\% and the uncertainties of the average values are small compared to this difference. Therefore, we included this difference in our systematic uncertainties.}

%-----Section 4-----%
\section{Results and Discussions}
\label{sec: result and discussions}
%-----Section 4-----%

In this section, we first present the results of theoretical calculations of the A-coefficients of the relevant transitions.
Then we present our final experimental results and compare them with theories.

\subsection{Theoretical calculation of A coefficients}\label{sec: theory}
In this work, relativistic calculations of Cs energy levels and transition rates were carried out using configuration-interaction many-body perturbation theory (CI+MBPT) \cite{CI-MBPT}, which has been successfully applied to many atoms and ions.
In this method, correlations among valence electrons are treated with the configuration-interaction (CI) method, while core-core and core-valence electron correlations are considered by the many-body perturbation (MBPT) theory.
In this work, we used the $V^{N-1}$ approximation, $i.e.$,  we took the Xe core and
considered a single valence electron. 
The basis set [25s25p25d25f25g25h] was employed, in which the 1-5s, 1-5p, 1-4d orbitals are treated as core orbitals, the 6-8s, 6-8p, 5-7d, 4f orbitals are treated as the valence orbitals, and the remaining orbitals are considered as virtual orbitals. 
The core and valence orbitals were obtained by the Dirac-Hartree-Fock calculation, and the virtual orbitals were constructed based on these core and valence orbitals. 
The CI+MBPT calculation was performed using this basis set in combination with one- and two-electron radial integrals, and then the electric and magnetic multipole transition moments including the random-phase approximation corrections were evaluated based on the calculated wavefunctions. 
\par
 The results of the calculations are listed in the 6th column of Table \ref{tab:results}. For allowed transitions, there also exist calculations performed by Safronova {\it et. al.} \cite{safronova}: they are listed in the 8th column.

Uncertainties in the calculated E2 transition rates is rather difficult to estimate, because we only performed the CI+MBPT level of calculation, and did not perform higher level calculation as was done in Safronova {\it et al.} for error estimation. The difference of $8\,^2P_{3/2} \rightarrow 5\,^2D_{5/2}$ E1 transition rate between our result and Safronova {\it et al.} is about 40 $\%$. If we can approximate that this difference is similar in the E2 transition, then the calculated result in the Table \ref{tab:results} may contain 40 $\%$ uncertainty.

\subsection{Results and comparison with theory}
Our experimental results are summarized in Table \ref{tab:results}.  The 4th column is the efficiency-corrected counting rate ($R$) proportional to the A-coefficient. The errors indicate only the statistical ones.
The 5th column is $R$ normalized to that of the allowed transition of $8\,^2P_{J}\rightarrow 5\,^2D_{3/2}$. 
 Here the second error indicates a systematic error: for the forbidden transitions both random and scale errors, listed in Table \ref{tab: uncertainty}, are added in quadrature while for the allowed transitions only the random type error is considered. Note that the scale error is relevant only to the different bands (700 vs 900 nm). 

Our main results are
\begin{eqnarray}
 &&\frac{A(8\,^2P_{3/2}\rightarrow 6\,^2P_{1/2})}{A(8\,^2P_{3/2}\rightarrow 5\,^2D_{3/2})}= (7.32\pm0.50\pm 1.14)\times 10^{-5}\hspace{5mm}
 \label{eq: main result 1} \\
  &&\frac{A(8\,^2P_{3/2}\rightarrow 6\,^2P_{3/2})}{A(8\,^2P_{3/2}\rightarrow 5\,^2D_{3/2})}= (6.13\pm0.56\pm 0.96)\times 10^{-5} \hspace{5mm}
 \label{eq: main result 2} \\
  &&\frac{A(8\,^2P_{1/2}\rightarrow 6\,^2P_{3/2})}{A(8\,^2P_{1/2}\rightarrow 5\,^2D_{3/2})}= (9.10\pm0.91\pm 1.42)\times 10^{-6}\hspace{5mm}
 \label{eq: main result 3} 
\end{eqnarray}
\par
We are now able to compare these results with theoretical expectations shown in the 7th column of Table \ref{tab:results}. 
Considering some uncertainties in the theoretical values, we conclude that the agreement between theory and experiment is satisfactory. 
In the case of the allowed transition $8\,^2P_{3/2}\rightarrow 5\,^2D_{5/2}$, our experimental value is consistent with two theoretical expectations within a 2$\sigma$ range.

\subsection{Forbidden transition $5\,^2D_{5/2}\rightarrow6\,^2S_{1/2}$}\label{sec: E25D6S}
In our setup, we also observed the transition $5\,^2D_{5/2}\rightarrow6\,^2S_{1/2}$ with a wavelength of 685 nm, depicted in Fig.\ref{fig:raw sepctra}(a) and (d). This is another forbidden transition, and its A-coefficient is already measured: {$A=22.22\pm 0.24$ Hz \cite{tojo6s5de2}}. To infer the experimental A-coefficient of this transition requires knowledge of its parent ($5\,^2D_{5/2}$) population, which in turn demands a rate equation simulation.
The result has uncertainty because the simulation needs all A-coefficients leading to $5\,^2D_{5/2}$ and some of them are poorly determined.
In addition, a two-photon ionization path may alter the population. 
Despite these restrictions, we carried out the simulation and found the A-coefficient to be {$22.2\pm0.60\pm3.45 $ Hz} when using the 893 nm transition as a reference. 
Considering various uncertainties, we conclude that our value is consistent with the reported value shown above: it enforces the reliability of our experiment.

%-----Section 5-----%
\section{Summary}
\label{sec:summary}
%-----Section 5-----%

In this article, we have presented an experiment that measured the Cs electric quadrupole (E2) transitions of $8\,^2P_{J}\rightarrow6\,^2P_{J'}$.
To this end, we employed laser-induced fluorescence spectroscopy: a continuous-wave laser beam in resonance with the $6\,^2S_{1/2}\rightarrow8\,^2P_{J}$ transition was used to excite Cs atoms in a heated vapor cell, and spontaneous emissions emitted to a right angle to the laser beam were detected with a photomultiplier tube after passing through a monochromator.  
We determined the ratio of the E2 transition rate to that of the electric dipole (E1) transition $8\,^2P_{J}\rightarrow5\,^2D_{3/2}$ from the same excited state. 
Our main results are shown in  Eq.(\ref{eq: main result 1})-Eq.(\ref{eq: main result 3}). 
\par
We also calculated the $8\,^2P_{J}\rightarrow6\,^2P_{J'}$ transition probabilities with configuration-interaction many-body perturbation theory (CI+MBPT) and compared them with the experimental values. 
The agreement between theory and experiment is found to be good. 
\par
This inherently weak transition serves as a crucial reference point for evaluating coherent amplification mechanism and comprehending potential background effects arising from other forbidden transitions \cite{Wang-Axion}. 
With this result, Axion and/or dark photon search is in progress. 

\begin{acknowledgments}
This work was supported by JSPS KAKENHI Grant Numbers 16H02136 (NS), 21H00074, 18K03621 24K07018 (MiT), 23K22520 (YM) and 21H01112 (HH). The computation in this work was partially performed at Research Center for Computational Science, Okazaki, Japan (Project: 23-IMS-C070).
\end{acknowledgments}

%-----Section 6-----%
\appendix
%\section{Appendixes}
%\label{appendix:qe}
%-----Section 6-----%
\begin{figure}[b]
   \centering
   \centering
  \includegraphics[width=\columnwidth]{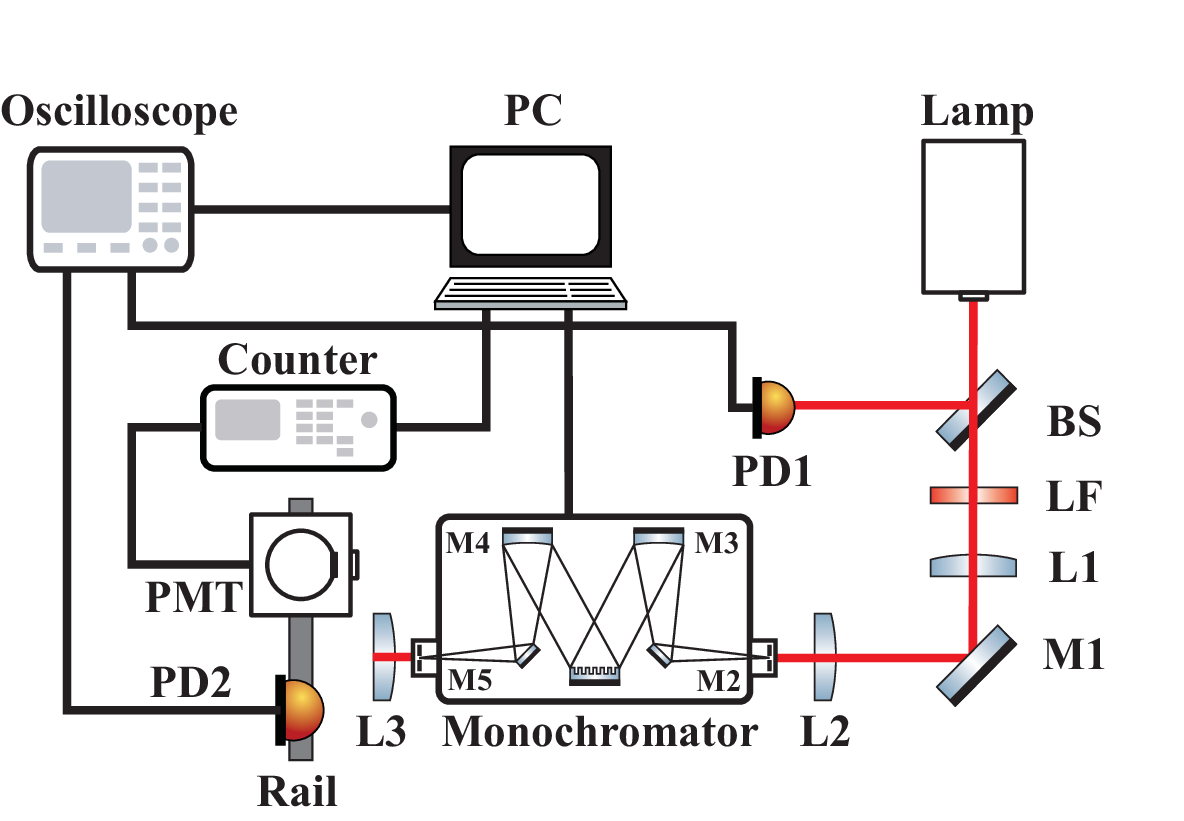}
   \caption{Calibration experiment setup. BS: beamsplitter, LF: lamp filter (FEL600 for QE calibration; FBH710-10 or FBH900-10 for $\eta_{opt}$ calibration, all from Thorlabs), other notations are consistent with those in Fig. \ref{fig:experiment setup}.}
   \label{fig:calibration setup}   
\end{figure}
\begin{figure}[b]
 \centering
 \includegraphics[width=\columnwidth]{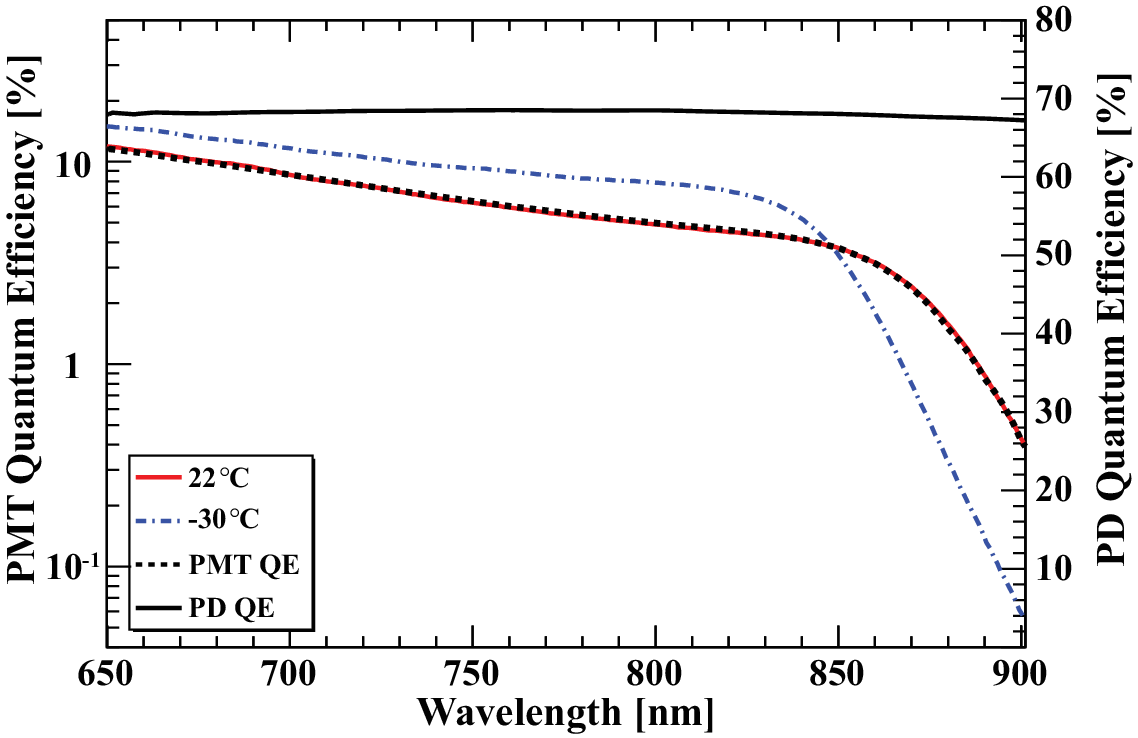}
 \caption{Quantum efficiencies vs wavelength. Our measurement: PMT QE at room temperature (red solid) and at -30 $^{\circ}$C (blue dash-dotted). Hamamatsu measurement: PMT QE at room temperature (black dashed line) and PD QE at room temperature (black solid).}
 \label{fig:pmttemp}
\end{figure}

\section{Photo-cathode quantum efficiency $\eta_{qe}$}
The photo-cathode quantum efficiency (QE) of PMT is one of the most important quantities since its differences for the signal ($684\sim715$ nm) and reference ($886\sim893$ nm) bands are expected to be large. We measured relative QEs experimentally and compared the results with those provided by the manufacturer. In this section, we describe our QE calibration procedure in detail.

For the QE calibration experiment, as shown in Fig.\ref{fig:calibration setup}, we replaced the Cs vapor cell with a Tungsten-Halogen lamp (Ocean optics, DH-2000), which emits a broad spectra ranging from 350 nm to at least 1 $\mu$m, with a power output of $>$1 W. The band pass filter (BP) was removed and a long pass filter (LF, Thorlabs FEL600) was placed before L1 to prevent higher order diffractions of the monochromator grating. We first adjusted the lamp intensity so that the PMT counting rate was less than $100$ kHz to avoid possible saturation effects.
We then measured the PMT rates by changing the monochromator wavelength from $590$ to $920$ nm at every 0.5 nm. 
The actual counts per point were greater than 400 Hz. Backgrounds, measured with the lamp blocked, were subtracted from the data. The procedure was repeated at different PMT temperatures: 22 $^{\circ}$C (room temperature), 15 $^{\circ}$C, -10 $^{\circ}$C, -20 $^{\circ}$C and -30 $^{\circ}$C.  
 We also measured a spectrum with a photo-diode (PD) (Hamamatsu, S2281) at room temperature by replacing PMT using a rail. Since the light intensity is rather weak for PD, so we use the serial amplifiers (Hamamatsu, C9051-01 and Thorlabs, AMP200) to measure the spectra. All PMT measurements are sandwiched by PD measurements to ensure the stability of the lamp spectra.
Note that the optics from the lamp to the detectors remained the same except for the neutral density filter (Thorlabs, ND10A) which was removed for PD to enhance the signal amplitude. 
Using the transmission spectra of ND10A, which was measured in a separate experiment, and the PD's QE provided by Hamamatsu \cite{S2281}, we could determine the lamp spectra at the exit of the monochromator, which was in turn used to determine PMT relative QEs.

Figure~\ref{fig:pmttemp} shows our results: the QE at room temperature (red solid), at -30 $^{\circ}$C (blue dash-dotted). Additionally, the QEs for the PD (black solid line), obtained from Hamamatsu's datasheet, and for the PMT (black dashed line), calibrated by the Hamamatsu factory, are also depicted. Note that our results are re-scaled so that the room temperature QE at 850 nm aligns with Hamamatsu's value. We find a good agreement between two room temperature QEs of ours and Hamamatsu's. The main uncertainties of the QE measurement come from the reproducibility of PD spectra, which is found to be better than 5\%.
\par
\begin{figure}[t]
\vspace{5mm}
 \centering
 \includegraphics[width=\columnwidth]{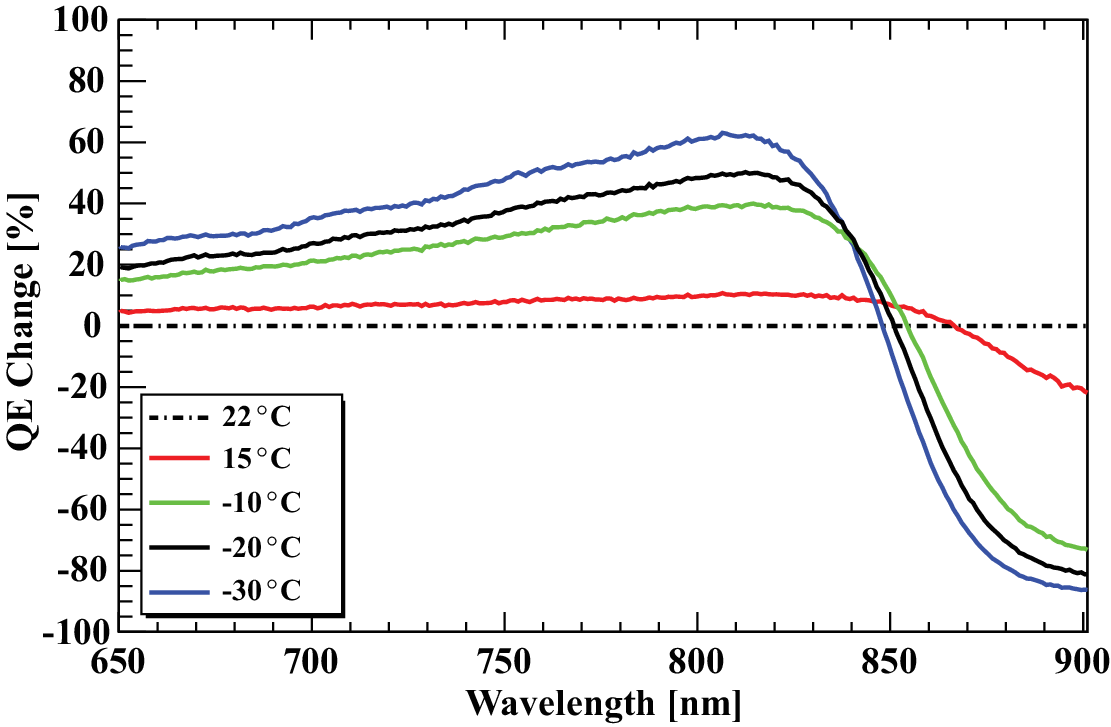}
 \caption{Quantum efficiency change normalized to room temperature, {\it i.e.}$\Delta \eta_{qe}/\eta_{qe}^{(0)}$.}
 \label{fig:qechange}
\end{figure}

Although not essential to our main experiment, we could measure the temperature coefficient of the cathode sensitivity as a byproduct.
The result is shown in Fig. \ref{fig:qechange}, where the vertical axis denotes $\displaystyle \frac{\Delta \eta_{qe}}{\eta_{qe}^{(0)}}=\frac{\eta_{qe}(T)-\eta_{qe}^{(0)}}{\eta_{qe}^{(0)}}$ with $\eta_{qe}^{(0)}$ being the room temperature QE.
 We note the temperature values used so far are all nominal (the values displayed by the cooling unit). We measured the actual temperature inside the cooling unit with a pt100 sensor. 
The measurement results are 15.6 $^{\circ}$C (15 $^{\circ}$C), -7.9 $^{\circ}$C (-10 $^{\circ}$C), -16.5 $^{\circ}$C (-20 $^{\circ}$C), and -26.5 $^{\circ}$C (-30 $^{\circ}$C), where the values in parentheses are nominal ones.\\[5mm]

\section{Optical component efficiency $\eta_{opt}$}
Another important quantity in Eq.(\ref{eq:raw_counting_rate}) is $\eta_{opt}$.  
It represents a product of efficiencies of all optical components from the Cs cell to the detector, including lenses (L1-L3), mirrors (M1-M5), BP and the grating.
In this section, a calibration experiment is described,  
 which measured 
 $\displaystyle r=\frac{\eta_{opt}(710\;{\rm nm})}{\eta_{opt}^{(0)}}$, 
the ratio of $\eta_{opt}$ at the signal band (684 $\sim$ 715 nm) and reference band (886 $\sim$ 893 nm).
As illustrated in Fig.\ref{fig:calibration setup}, we inserted lamp band pass filters (LF) with the transmission window of $710 \pm 5$ nm or $900\pm 5$ nm at the exit of the lamp, and measured the intensity with PD right after the L1 and the exit of the monochromator. There was no BP in this setup.

The PD output after the LF of $\lambda_{LF}=$ (710 nm or 900 nm) is proportional to $I(\lambda_{LF}) \eta_{PD}(\lambda_{LF})$ while that at the monochromator exit is proportional to $I(\lambda_{LF}) \eta_{PD}(\lambda_{LF}) \eta_{opt}(\lambda_{LF})$, where $I$ and  $\eta_{PD}$ denote the lamp intensity and the PD sensitivity, respectively. 
We deduced $r$ from these 4 measurements (two different LF at two different locations) and found it to be $r=1.54$. Three corrections are applied to $r$ obtained above to reach the values in Table \ref{tab:raw counts}. 
(i) The effect of the BP (as shown in Table \ref{tab:scanned-wavelength}): we measured the transmission efficiency of each combination in a separate experiment. (ii) The L1 lens effect: since the PD detector is placed downstream of L1 in the calibration setup, we must take into account the wavelength dependence of its transparency separately. Fortunately, the lens material (BK7), has flat transparency in this wavelength region: 92\% at 710 nm and 94\% at 900 nm (10 mm thick). (iii) Wavelength dependence within the signal or reference bands. The main dependence comes from the grating efficiency. With the efficiency curve provided by the manufacturer \cite{SP2300i}, corrections are estimated. Uncertainty in $\eta_{opt}$ comes from the non-reproducibility of PD spectra. Considering the results of $\eta_{qe}$  calibration experiments, we assign 5\% uncertainty to $\eta_{opt}$ conservatively. 

%-------bibliography--------%
%apsrev4-2.bst 2019-01-14 (MD) hand-edited version of apsrev4-1.bst
%Control: key (0)
%Control: author (8) initials jnrlst
%Control: editor formatted (1) identically to author
%Control: production of article title (0) allowed
%Control: page (0) single
%Control: year (1) truncated
%Control: production of eprint (0) enabled
%

\end{document}